 \documentclass[final,3p,times]{elsarticle}

\usepackage{amssymb}
\usepackage{amsmath}
\usepackage[bookmarks=false]{hyperref}

\usepackage{graphicx}

\usepackage{multirow}

\DeclareMathOperator*{\argmin}{arg\,min}

\journal{Biomedical Signal Processing and Control}

\begin{document}

\newcommand{\orcid}[1]{\href{https://orcid.org/#1}{\includegraphics*[width=10pt]{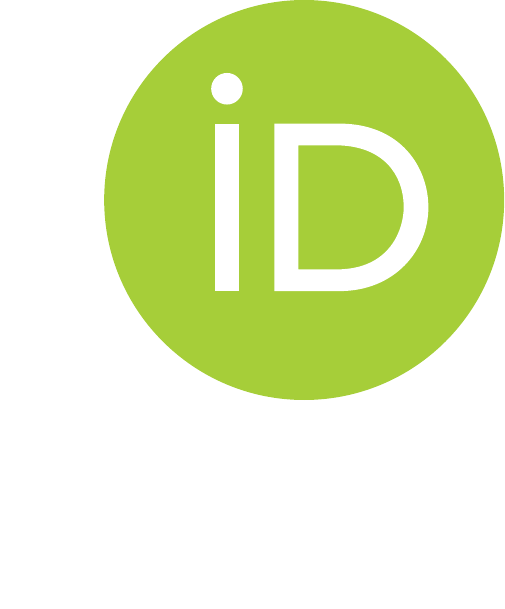}}}

\begin{frontmatter}



\title{Mathematical model of parameters relevance in adaptive level-crossing sampling for electrocardiogram signals}


\author[1]{Silvio~Zanoli\orcid{0000-0002-0316-1657}}
\author[1]{Giovanni~Ansaloni\orcid{0000-0002-8940-3775}}
\author[2]{Tomás~Teijeiro\orcid{0000-0002-2175-7382}}
\author[1]{David~Atienza\orcid{0000-0001-9536-4947}}
\address[1]{Embedded Systems Laboratory (ESL), École Polytechnique Fédérale de Lausanne (EPFL), 1015 Lausanne}
\address[2]{Basque Center for Applied Mathematics (BCAM), Bilbao, Spain}


\begin{abstract}
Digital acquisition of bio-signals has been mostly dominated by uniform time sampling following the Nyquist theorem. However, in recent years, new approaches have emerged, focused on sampling a signal only when certain events happen. Currently, the most prominent of these approaches is Level Crossing (LC) sampling. 
Conventional level crossing analog-to-digital converters (LC-ADC) are often designed to make use of statically defined and uniformly spaced levels. However, a different positioning of the levels, optimized for bio-signals monitoring, can potentially lead to better performing solutions.
In this work, we compare multiple LC-level definitions, including statically defined (uniform and logarithmic) configurations and optimization-driven designs (randomized and Bayesian optimization), assessing their ability to maintain signal fidelity while minimizing the sampling rate. In this paper, we analyze the performance of these different methodologies, which is evaluated using the root mean square error (RMSE), the sampling reduction factor (SRF) ---a metric evaluating the sampling compression ratio---, and error per event metrics to gauge the trade-offs between signal fidelity and data compression. Our findings reveal that optimization-driven LC-sampling, particularly those using Bayesian methods, achieve a lower RMSE without substantially impacting the error per event compared to static configurations, but at the cost of an increase in the sampling rate.
\end{abstract}



\begin{keyword}
Non-uniform sampling \sep Bio-signal monitoring \sep Event-based \sep Level-Crossing \sep ECG \sep Bayesian optimization \sep Level-Crossing optimization
\end{keyword}

\end{frontmatter}
\section{Introduction}
\label{sec:intro}

Today, chronic diseases and, in particular, chronic heart disease are the leading cause of death worldwide~\cite{cardio_diseases_leading_death_cause}.
The monitoring of such conditions is often challenging and not achievable in hospital environments, as it requires long-term continuous bio-signals recording in normal day-to-day activities. Such long-term observations call for non-invasive solutions, making wearable sensors a key technology for such applications.

The energy budget of any portable device can be divided into four categories: computation, storage, communication, and data acquisition. Decades of research have culminated in highly optimized systems when it comes to computation, storage, and communications~\cite{Wei19, Abadal20, Panades20, Pullini18}. However, the energy budget of modern wearable systems is highly affected by the signal acquisition component~\cite{rincon2011development}. Hence, more efficient signal sampling strategies have the potential to greatly increase the efficiency of battery-operated devices.

The usual approach to this problem is the uniform time sampling, first mathematically described in their foundational work by Nyquist and Shannon. In their analysis, they established a lower limit to the sampling rate, known as the Nyquist frequency, which is twice the maximum frequency present in the signal under analysis~\cite{Shannon1949}. Such an approach ensures, in the ideal case, a perfect signal reconstruction (stable and unique). However, when the assumption of a stable and unique signal reconstruction is removed, different sampling methodologies can be used to represent the data. One prominent family of non-uniform sub-Nyquist sampling methods is event-based (EB) sampling, a data-acquisition strategy designed to capture signal points only when specific events occur within the signal.
Such methodologies aim to acquire samples depending on the signal behavior. Although these methods are often lossy and do not guarantee perfect signal reconstruction, they often trade the resulting error for a reduced number of acquired samples and energy efficiency. When the discarded signal sections are not significant for the task at hand, such methodologies allow for far grater device autonomy and efficiency. 

The most common technique in EB-signal acquisition is level-crossing  ~\cite{level_crossing}, where an event is generated whenever the measured physical quantity crosses predefined levels. This can be done in two ways: Analog level-crossing~\cite{analog_level_crossing}, where a sampler detects crossings of analog-defined levels, and Digital level-crossing~\cite{level_crossing}, where a custom logic defines digital levels, the signal is first digitized using an analog-to-digital converter (ADC) and samples are forwarded to the processing unit only when the digital levels are crossed.

These methods have demonstrated the ability to significantly reduce the average sampling frequency of real-world signals by up to 90\%~\cite{polygonal_approximator}. However, as the reduction rate increases, the fidelity of the signal significantly deteriorates, as these acquisition methods are inherently lossy. Moreover, such sampling schemes are often highly parametric. In fact, a Level Crossing (LC) ADC is functionally characterized by the number of levels, their position, and, if present, the level hysteresis value. 

The focus of this work is to study the effect and optimality of such parameters in LC sampling schemes. To this end, in the following sections, we explore several solutions to the level definition problem and highlight their strengths and weaknesses by showcasing the effects on signal reconstruction accuracy for different sets of level values. In particular, we target four methodologies:
\begin{itemize}
    \item Uniform levels distance
    \item Logarithmic distanced levels
    \item Random search optimized levels
    \item Bayesian optimized levels
\end{itemize}

\begin{figure}[ht]
    \centering
    \includegraphics[width=\columnwidth]{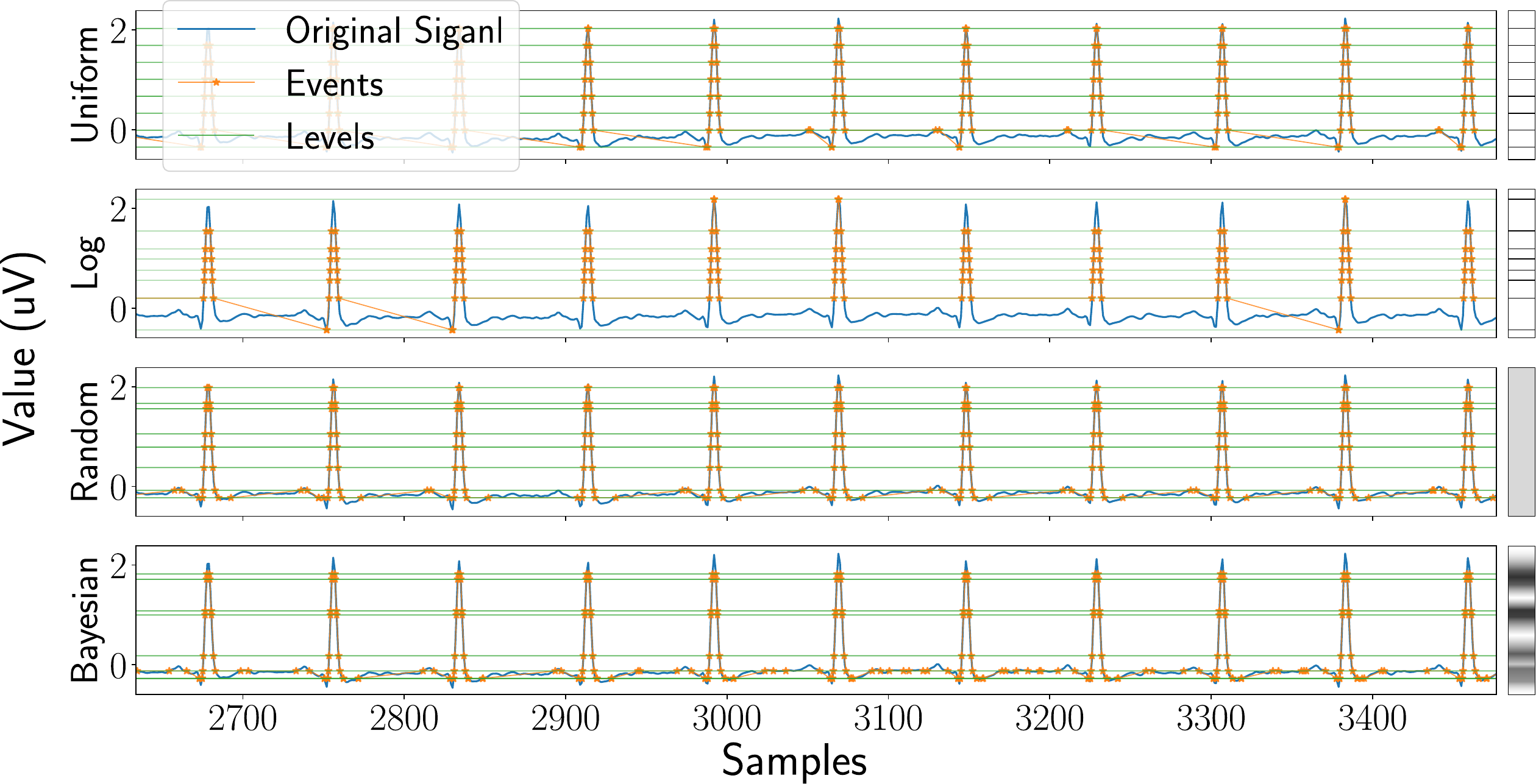}
    \caption{Illustration of different methodologies for level selection. The rectangles on the right of each graph show the uncertainty of the associated algorithm when choosing a level value.}
    \label{fig:methods}
\end{figure}

Figure~\ref{fig:methods} shows the intuition behind each method. Uniform levels consist of equally spaced levels that span the range of the signal under analysis.
Logarithmic levels are placed on a symmetric logarithmic scale centered on zero, similarly to what is proposed in \cite{log_lvls}.
These first two methods are the most widely used and readily available level definition techniques, as they are statically defined and do not require further signal analysis.
On the other hand, optimization-based methods have the potential to improve on a designated merit figure (e.g. the reconstruction error). However, the sampling problem is rarely differentiable, especially in the context of EB-sampling. To assess this problem, random search is a gradient-free optimization methodology that does not rely on prior knowledge and selects the best result out of a series of randomly sampled solutions. Moreover, Bayesian optimization is another gradient-free optimization approach, where each successive optimal guess is the solution of the optimization problem of a surrogate model, conditioned with every subsequent guess.

For each methodology, we analyze the root mean square error (RMSE) between the original signal and the linear resampled events recorded. We also show the compression ratio obtained by each methodology by comparing the number of events acquired with the number of samples in the original signal. Finally, we also present the RMSE normalized relative to the number of events, giving a measure of error per recorded event. 

To this end, we created an implementation of an LC-ADC, optimized for fast execution in a functional manner, using the JAX library\cite{jax2018github}. The details, together with all the rest of the code that allows the reproduction of our results, can be found in the repository in \footnote{eslgit.epfl.ch/esl/algorithms/biosignal-processing/lc\_learning}.

Finally, since Bayesian optimization is a methodology rich in meta-parameters, we are interested in showing how they can affect the optimization results. We thus perform a sensitivity analysis of such meta-parameters using a least-square linear model fit.

The techniques we analyze in this work do not dependent on the type of the target bio-signal. As a specific experimental use case to test our methodology, we consider long-term electrocardiograms (ECG). Such signals are often acquired over long periods of time (a day or more) with low-power wearable devices. These types of applications, were data acquisition is a long and continuous process, are the prime candidates for EB-sampling based techniques as they sensibly reduce the amount of energy required by the whole system.


The manuscript proceeds as follows. Section \ref{sec:SOA} illustrate state of the art heuristic techniques used for levels placement in different applications. Section \ref{sec:methods} then shows how a gradient-based approach is not feasible in this domain, introducing two gradient-free approaches and two statically defined levels methods, used as baselines. Section \ref{sec:experimental_setup} sets the empirical methods used to evaluate our work. Finally, Section~\ref{sec:results} shows the results obtained, highlighting the strengths and weaknesses of optimization-based methodologies.

\section{State of the art}
\label{sec:SOA}

The topic of signal sampling has been extensively explored in the last eighty years. Starting from the seminal work of Shannon and Nyquist \cite{Shannon1949} with time-centered signal acquisition techniques. Other approaches~\cite{non_uniform_sampling_reconstruction_conditions,sub_nyquist_optimal} broaden this condition while still achieving a loss-less reconstruction of the sampled function. 

Many EB-schemes exist, LC being the most prominently used but not the only one explored in the literature. One notable example of such an EB sampler without LC is the polygonal approximation method~\cite{polygonal_approximator}, also called the integral criterion method, in which the signal is first digitized with an ADC, but samples are only sent to the main processing unit when the error between the uniformly sampled signal and its linear approximation exceeds a certain threshold $\varepsilon$.

Moreover, of relevance is the relationship between EB-sampling and Compressed Sensing (CS)~\cite{compressed_sensing}, which compresses data using signal-agnostic measurement matrices to create a sparse representation. Unlike EB-sampling, the length of this sparse representation in CS is independent of the signal.

Focusing on LC-sampling schemes: in \cite{LC_lvls_in_SRF_function} and in \cite{Average_Samp_freq_LC}, the authors discuss the data rate resulting from the LC-sampling of signals with known statistical properties. 
This analysis is then explored in \cite{EB-PAS-analysis}, discussing the required bandwidth to send data acquired following the EB-sampling integral criterion.
This topic is then further expanded in \cite{EB_for_control}, where EB sampling is analyzed in a control-theory framework and the mathematical formulation of the average sampling frequency is extended to send-on-area, send-on-energy, and Lyapunov sampling.
The work in \cite{uniform_lvls_control} analyzes the effects of LC acquisition when integrated into a control loop system, highlighting that when compared to classical sampling, it gives substantially smaller variances in the system response for the same average sampling rates.
Then, in \cite{many_EB-samp_for_control} the authors extended the analysis of EB-sampling in control systems by empirically simulating four different types of sampling schemes and highlighting that a significant reduction of the event rates could be achieved with small or negligible deterioration of the system performance.

Although all of the above methods rely on statically and uniformly separated levels, the work in~\cite{pdf_based_LCS} derives a law based on $\mu$ (based on average) to define the positioning of levels in signals with a known probability density function. However, the solution found is non-optimal and does not tackle the problem of hysteresis.
Similarly, in \cite{adaptive_LC_recon} the authors use local averages to adaptively compute average-based levels. Then, they used the events and the notion that they have been recorded in signal averages to drive the signal reconstruction through prolate spheroidal wave functions.
In \cite{energy_intensive_cool_adaptive_LC}, a circuit for adaptive resolution LC-ADC is presented. This approach increases or decreases the number of levels used according to the local signal reconstruction error. Although this technique achieves outstanding performance, it requires additional custom hardware to condition an LC-ADC that must be designed to accommodate a variable number of levels. Moreover, the error evaluation method must continuously compare the recorded events with the underlying signal, greatly decreasing the potential energy savings.

In contrast, the present work aims at comparing several LC-based sampling schemes and analyzing the resulting level-hysteresis effect and optimal value in long-term energy-efficient bio-signal acquisition. The methodologies choice and evaluation is experiment-driven with attention on the underlying mathematical problem, driving the choice of levels position optimizers. We show that statically defined levels, optimized a priori on a subset of the ECG signal, can greatly outperform statically uniformly defined levels.





\section{Methods}
\label{sec:methods}
In this Section we discuss the aforementioned levels definition methodologies, defining their functional behavior.

First, in Section \ref{sub-sec:model} we analyze a formal model for the LC sampling problem, highlighting the need for a gradient-free approach to parameter optimization. In Section \ref{sub-sec:param_selec_methods}, we show the different methodologies we selected, highlighting their strengths, weaknesses, and reasons for the choice. Finally, while most of the level-selection methods analyzed are self-reliant and data-agnostic, the presented Bayesian optimization method is both a gradient-free optimization methodology and highly parametric. These characteristics lead to optimal results when compared to other methodologies, with the added costs of high results variance as function of the algorithm parameters. Hence, Section \ref{sub-sec:ANOVA_description} proposes a sensitivity analysis design based on a linear model with interaction.

\subsection{Level-crossing model}
\label{sub-sec:model}
Given a signal $x(t)$, with $t$ the time in which the signal evolves, we can write the general sampled form as 
\begin{equation}
    \label{eq:generic-samp}
    S(t,\theta) = x(t)\delta(g(t,\theta))
\end{equation}
Where $S(t,\theta)$ is the time-continuous (i.e. with continuous support) sampled signal, $\delta(\cdot)$ is the Dirac impulse function, $g(t,\theta)$ is a sampling function that is 0 only in the sampling instants, and $\theta$ is the vector of sampling parameters, such as time intervals, levels, maximum integral error or others.
For $g(t,\theta) = N_{y}(t,T)= \prod_{i=-\infty}^{\infty}(t-iT)$ we obtain the standard uniform-time sampling with period $T$.
The common objective when sampling any physical quantity is the ability to recover the original signal as closely as possible from its samples.
To do so, we make use of an interpolator function that in the continuous-time sampling representation takes the form of any function that given $S(t,\theta)$ returns an approximation of $x(t)$ on the same support (e.g. linear interpolation between any two consecutive samples). In particular, we write the interpolator function as  $I(S(t,\theta),\varphi)$, with $\varphi$ the interpolator parameters (e.g. the 0-crossing and period and scale of a sinc function). We are now interested in finding $\theta^{*}$ that minimize the difference between $x(t)$ and the resampled signal, hence:
\begin{equation}
    \label{eq:error}
    \theta^{*} = \argmin_{\theta}\int_{-\infty}^{\infty}|I(S(t,\theta),\varphi)-x(t)| dt
\end{equation}
In general, we can not assume $\nabla_{\theta}S(t,\theta)$ to be well defined, and therefore a gradient-based approach parameter optimization is rarely possible. Moreover, such an approach would require a predefined interpolator function which cannot be proven optimal without further functional analysis of the sampling scheme, making the problem a mixed functional optimization task. Although some special cases admit an analytical solution, such results are not obtainable through gradient descent methods. For example, we can notice how the Fourier transform of $S(t,\theta)$ in band-limited and uniformly sampled $x(t)$ is composed of a spectral replica of $x(t)$, this leads to the $Sinc$ interpolator with $\theta^* = \varphi = T^*  \geq \frac{1}{2 \cdot BW}$, where $BW$ is the $x(t)$ bandwidth ~\cite{Shannon1949}.

However, this approach is not always possible. Consider now 
\begin{equation}
    \label{eq:LC_eq}
    g(t,\theta) = \prod_{i=0}^{K-1}(x(t)-\theta_l(i))
\end{equation}
With $\theta_l(i)$ being a set of $K$ real numbers.
Such sampling function is the level-crossing sampler, with $K$ levels, defined by $\theta_l(i)$, and without hysteresis. This leads to a $S(t,\theta)$ without any clear properties, even when analyzed in the Fourier domain. If we then include the hysteresis term, we can see how this approach would become quickly intractable due to its  non-linear nature.

We hence propose to analyze a set of gradient-free optimization methodologies when the interpolator function $I(\cdot)$ is a linear interpolation between samples, and compare such techniques with statically defined alternatives.

\subsection{Level-crossing parameter selection methods}
\label{sub-sec:param_selec_methods}
As we discussed in Section \ref{sec:SOA}, several methodologies can be applied when deciding the parameters of LC sampling. In this work, we first analyze the two most common approaches: uniformly spaced and logarithmically spaced levels. Then, we introduce two new techniques aimed at optimizing the reconstruction error, namely, Random search and Bayesian-optimized parameters. These last techniques were chosen because of their ability to find a locally optimal solution to gradient-free problems.
Such techniques allow us to obtain a set of locally-optimal LC parameter with a top-down approach: designing an objective function that captures the reconstruction error and finds the parameters that minimize it.

In Figure \ref{fig:methods} we illustrate the results obtained by each of them on an example ECG signal. We can notice how the statically defined methodologies do not have any uncertainty relative to the levels positioning as they are decided beforehand. The random optimization does not use any prior information to draw the next set of levels, and hence its uncertainty is uniform across the range. The Bayesian optimization method manages to gradually reduce this uncertainty as more configurations are tested.

When working with digital systems, sampled signals cannot be represented as a train of Dirac pulses as seen in Section \ref{sub-sec:model}, therefore, we represent it with a sequence of signal values at the corresponding sampling instants. Given a physical signal $x(t)$ and its uniform-time sampled version with interval $T$: $x[i]=x(i\cdot T)$, we obtain a list of events obtained emulating the LC sampling process. We then use a linear interpolator between each event (as we have shown in \cite{WFDB_zanoli} to be the best performing common interpolator for the LC-sampled ECG signals) obtaining $x^*(t)$. Finally, we uniform-time resample such signal with interval $T$: $x^*[i]=x^*(i\cdot T)$ to obtain the values of the LC-interpolated signal at the same timing instants as $x[i]$. 

The objective function for optimization is hence defined as:
\begin{equation}
    \label{eq:obj}
    E_{rr} = \sqrt{\frac{\sum_{i=0}^{N}|x[i]-x^*[i]|^2}{N}}\cdot(1+\lambda \cdot SRF)
\end{equation}
Where $x[i]$ is the $i^{th}$ point of the uniformly sampled signal, $x^*[i]$ is the $i^{th}$ point of the resampled event-based sampled signal, $N$ is the length of the signal, $\lambda$ is a regularizer: $\lambda\in\mathbb{R}$ and $SRF$ is the sampling reduction factor, defined as 
\begin{equation}
    \label{eq:SRF}
    SRF = \frac{number~of~events}{number~of~uniform~samples}
\end{equation}
This last regularization factor is included as we are interested in observing the trade-off between compression and signal representation. However, since the methodologies defining static levels do not use any regularizer parameter, as they do not perform any optimization, using this term in the optimization would lead to an unfair comparison between RMSE and a regularized version of this metric. Hence, experiments that compare the different methodologies will show the results obtained for $\lambda=0$. However, in the optimization meta-parameters sensitivity analysis, shown in Section \ref{sub-sec:ANOVA_description}, we will investigate the effect of such a parameter.

Given an LC sampling scheme with $L$ levels $\theta_i$ (with $i \in \{0,...,L-1\},~L>1$), applied to a signal that spans between a maximum $x_{max}$ and a minimum $x_{min}$, defined as the $5^{th}$ and the $95^{th}$ percentile values of the signal, and with range $x_{max}-x_{min} = \Delta$ we obtain the levels for the different methodologies as follows:

\textbf{Uniformly spaced levels}: Each level $\theta_i$ follows the formula:
\begin{equation}
    \theta_i = x_{min} + \frac{\Delta}{L-1}\cdot i,~i\in\{0,..,K-1\}
\end{equation}
Which is a uniform partition of the range $\Delta$ into $L-1$ intervals with the levels being positioned at the interval edges.

\textbf{Logarithmically spaced levels}: The levels definition methodology presented here is relative to the LC with an odd number of levels. To obtain the values for an even $L$ we use the same definition, computing $L+1$ levels and removing the one positioned on 0. To obtain the discussed symmetric logarithmic spacing centered around 0 we first generate a log space between 1 and 10 with $\lceil\frac{L}{2}\rceil$ values. We invert such a scale and shift it by two times the smallest value obtained (which is 1, so we just add 2) obtaining two specular logarithmic sets that start from 1. We then concatenate these two sets of values, removing one of the duplicate central levels, and scaling it to fit the range of the signal. Such operations are formally defined as follows:

\begin{equation}
    \theta^{+}_i = 10^{\frac{i}{\lfloor\frac{L}{2}\rfloor}},~i\in\{0,..,\lfloor\frac{L}{2}\rfloor\}
\end{equation}

\begin{equation}
    \theta^{-}_i = -\theta^{+}_{i}+2,~i\in\{1,..,\lfloor\frac{L}{2}\rfloor\}
\end{equation}

\begin{equation}
    \theta_i = Norm(Sort(\theta^{+},\theta^{-}))_i \cdot \Delta + x_{min},~i\in\{0,..,K-1\}
\end{equation}

Where $\theta^{+}$ and $\theta^{-}$ are the two described positive and negative log levels, $Sort(\cdot)$ is the function that returns the sorted list of numbers for a given sequence and $Norm(\cdot)$ is the Min-Max normalization, transforming the input list between $0$ and $1$.

\textbf{Random search (RS)} is a gradient free techniques particularly suitable for optimizing functions that are non-continuous or non-differentiable. This methodology explores the search space by random sampling unique points in it and keeping the best performing ones, allowing it to handle complex objective functions with unknown or discontinuous structures. This is especially useful in the context of meta-parameter tuning in level crossing signal acquisition, where traditional gradient-based methods can not be used. Our implementation consists of a random number generator, producing a set of unique random number with length $N = \alpha \cdot L$, with $\alpha$ the number of random LC schemes tested (in our experiments, $10000$), $L$ the number of levels. Then, we subdivide this set into subsets $\alpha$ to obtain the level values to test. Finally, we repeat the same operation to obtain one hysteresis value for each LC scheme found.  

Formally:
\begin{equation}
    \Theta = Norm(rand_{nr}(\alpha \cdot L,\mathbb{N}))\cdot\Delta+x_{min}
\end{equation}
\begin{equation}
    \Theta^{adc}_i =  \{\Theta[i\cdot L],...,\Theta[i\cdot L+L]\}, i\in \{0,...,\alpha-1\}
\end{equation}

\begin{equation}
    \Gamma = Norm(rand_{nr}(\alpha ,\mathbb{N}))\cdot hist_{max}
\end{equation}
\begin{equation}
    \gamma_i = \Gamma[i], i\in \{0,...,\alpha-1\}
\end{equation}
Where $rand_{nr}(\cdot,\cdot)$ is a random generator, choosing a set of non repeating number from the set of natural numbers $\mathbb{N}$, $\Theta$ is the list of all levels for all random LC sampler, $\Theta^{adc}_i$ is the $i^{th}$ set of random levels, $\Gamma$ is the set of all hysteresis, and $\gamma_i$ is the hysteresis of the $i^{th}$ LC sampling solution. Finally, $hist_{max}$ is the maximum allowed hysteresis value.

\textbf{Bayesian optimization (BO)} 
BO is an iterative, two-step optimization method. A detailed description of its functioning can be found in \cite{Byesian_opt_book}, but we can summarize it as follows:

Given a signal $x(t)$, its sampled form $S(t,\theta)$, and the interpolated reconstructed signal $I(S(t,\theta),\varphi)$, we are interested in finding $\theta^{*}$, as in eq.\ref{eq:error}. Since, in this work, the interpolator is assumed to be a linear resampler without parameters, and $x(t)$ is considered a known signal, this can be simplified in the objective function:

\begin{equation}
    \theta^*= \argmin_{\theta} E_{rr}(\theta)
\end{equation}
Where $E_{rr}$ is the objective function in eq.\ref{eq:obj}.

The Bayesian optimization method assumes the objective function to be a random process and builds a surrogate model of it, often using a Gaussian process prior.
\begin{equation}
    p(E_{rr}) = \mathcal{GP}(E_{rr};\mu,\sigma)
\end{equation}
Given the initial observations $\mathcal{D}=(\Theta,\mathbf{E_{rr}})$, where $\Theta$ is a random initial set of different $\theta$ (levels and hysteresis in this context), and 
$\mathbf{E_{rr}}$ are the relative measured errors, we condition the distribution to obtain

\begin{equation}
    \label{eq:gauss_proc_codn}
    p(E_{rr}|\mathcal{D}) = \mathcal{GP}(E_{rr};\mu_{E_{rr}|\mathcal{D}},\sigma_{E_{rr}|\mathcal{D}})
\end{equation}

Then, we use a sampling function to obtain the next point to test in the estimated random process. Many sampling functions have been proposed in recent years\cite{bayesian_opt_review}. We show here the one chosen for this work: the Expected improvement\cite{Byesian_opt_book}. Given a utility function
\begin{equation}
    \label{eq:utility_f}
    u(\theta) = max(0,E_{rr}'-E_{rr}(\theta))
\end{equation}
Where $E_{rr}'$ is the minimum value of $E_{rr}(\theta)$ observed so far, we compute
\begin{equation}
    \label{eq:acq_func}
    a_{EI}(\theta) = \mathbb{E}[u(\theta)|\theta,\mathcal{D}]
\end{equation}
We then select the point with the highest expected improvement $a_{EI}$ and measure $E_{rr}(\theta)$ in that point. This new information is then added to the prior $\mathcal{D}$, and the process is repeated until $a_{EI}$ falls below a certain tolerance or for a fixed number of iterations. The final $\theta$ obtained by this process is the locally-optimal value.

Implementations of the Bayesian optimization process are rich in parameters. These include the number of initial observations used to obtain the starting conditioning of the underlying Gaussian process, the minimal improvement to consider when maximizing the utility function (this is to avoid sampling points of negligible expected improvement that would be overshadowed by the process variance), and the maximum number of the algorithm iterations shall it not achieve convergence. In the next Section, we discuss the effects of such parameters when applied in the context of optimal level crossing.

\subsection{Bayesian optimization sensitivity analysis}
\label{sub-sec:ANOVA_description}

The Bayesian optimization implementation we use in this work needs the definition of four parameters. Moreover, following in Eq.~\ref{eq:obj}, the $\lambda$ regularization factor have an impact on the optimization results, as it regulate the importance given to the sub-sampling effects of LC sampling. To determine the general behavior of the Bayesian optimization algorithm in the current context, we compute the sensitivity of its factors by analyzing the fit of a linear mixed model to its parameters:

\textbf{Initially sampled points}: This parameter defines the number of random sampled LC sampler the Bayesian optimization methodology uses to initially condition the Gaussian process model of the optimized function.

\textbf{Optimally sampled points}: The number of iterations of the Bayesian optimization process, where an iteration is the process of sampling the optimal estimate from Eq.~\ref{eq:acq_func}, obtaining the true $E_{rr}$ value and condition the Gaussian process as in eq.~\ref{eq:gauss_proc_codn}.

\textbf{Minimum improvement ($\xi$)}: Defines a threshold for what constitutes an improvement over the previous optimal point in Eq.~\ref{eq:acq_func}. 

\textbf{Train length}: When evaluating the results of the optimization, we use a standard separation of the dataset in the training and test sets. For each ECG record, we define the length of the training set, measured in heartbeats, and consider the rest of the signal as test.

\textbf{Lambda ($\lambda$)}: The objective equation in \ref{eq:obj}, makes use of a regularizing $\lambda$. This parameter is a penalty given to the objective function that increases the error as more events are acquired. Although fixed to 0 when comparing the LC sampling optimization methods with the statically defined ones, we are interested to observe if such parameters could lead to beneficial results.

The effects these parameters have on the algorithm results are a function of both the signal reconstruction error and the number of events recorded. The used metric is hence the  root mean square error between the reconstructed and original signals, normalized on the relative number of recorded events, giving the average error per event:
\begin{equation}
    \label{eq:norm_error}
     Error_{norm} = SRF\cdot\sqrt{\frac{\sum_{i=0}^{N}|x[i]-x^*[i]|^2}{N}}
\end{equation}

Given a small enough domain, it is reasonable to fit a standard coefficient linear model with interactions to the function of the aforementioned meta-parameters, hence, the sensitivity analysis is the result of the best fit to the equation:

\begin{equation}
Error_{norm}(\beta) = K_0 + \sum_{i=1}^{|\beta|} K_{\beta_i} \beta_i + \sum_{i=1}^{|\beta-1|}\sum_{j=i+1}^{|\beta|} K_{\beta_i \beta_j} \beta_i \beta_j
\label{eq:linear_model}
\end{equation}


Where $K_0$ is the constant coefficient, representing the model independence from its meta-parameters, \\$\beta = \langle I_p,S_p,\xi,T_l,\lambda \rangle$ is the list of meta-parameters listed above in this Section in order of mentioning, standardized with respect to the analyzed range of each variable to the $[-1,1]$ interval, $K_{\beta_i}$ are the model coefficients:\\ $K_{I_p}, K_{S_p},K_{\xi},K_{T_l},K_{\lambda}$ to the corresponding main effects, and $K_{\beta_i \beta_j}$ are the coefficient of interactions between two meta-parameters: $K_{I_p S_p}$, $K_{I_p \xi}$, $K_{I_p T_l}$, $K_{I_p \lambda}$, $K_{S_p \xi}$, $K_{S_p T_l}$, $K_{S_p \lambda}$, $K_{\xi T_l}$, $K_{\xi \lambda}$, $K_{T_l \lambda}$.


\section{Experimental setup}
\label{sec:experimental_setup}
We design an experimental flow to characterize the behavior of the ECG signal with the different LC solutions produced by the methodologies described in Section \ref{sub-sec:param_selec_methods}. Section~\ref{sub-sec:dataset} specifies the dataset used for this study, the pre-processing applied to the records, and the assumption used in our work. Then, Section~\ref{sub-sec:setup} presents the model of the emulated LC acquisition used for this study, showing the operational pipeline used to obtain the LC sampling error with respect to the dataset signal, given the parameters obtained by the analyzed methodologies. Finally, Section~\ref{sub-sec:ANOVA_setup} defines the sensitivity analysis methodology of the Bayesian optimization method and its meta-parameters boundaries.

\subsection{Dataset}
\label{sub-sec:dataset}
Event-based sampling is of particular interest when applied to long-term patient monitoring. Hence, a key application for this technology is the recording of whole-day ECG signals. Moreover, the ECG signal is characterized by self-similar repeating elements~\cite{fractal_ECG_reconstruction}; such a characteristic is of particular interest for methodologies that take advantage of the data structure such as the proposed Bayesian optimization. The emulated process described in Section \ref{sub-sec:setup} makes use of previously recorded data to obtain the LC scheme performances. The dataset used in this work is the MIT-BIH Normal Sinus Rhythm database \cite{physioToolkit}. This database comprises 18 long-term (approximately one day) ECG recordings of subjects referred to the Arrhythmia Laboratory at Beth Israel Hospital in Boston (now the Beth Israel Deaconess Medical Center). The total number of heartbeats analyzed is approximately 1.6 million, or 90'000 per patient. The ECG recordings are sampled at a frequency of 128 Hz.

\subsection{LC sampling evaluation}
\label{sub-sec:setup}
The implemented experimental framework is composed of four components:
\begin{enumerate}
    \item Digital signal filtering
    \item Train-Test subdivision
    \item LC-ADC emulation with linear resampling assumption
    \item Error estimation between LC sampled signals and the original dataset
\end{enumerate}

\textbf{Digital signal filtering:}
The dataset described in Section \ref{sub-sec:dataset} is provided without detailed information on the acquisition process. However, an exploratory analysis of the recording shows that the dataset is probably unfiltered in its higher frequency components. Given the limited band in which an ECG signal is bounded to fall in \cite{HR_BW} a band-pass digital filter reduces the bandwidth of each record, using a Finite Impulse Response (FIR) filter design using the least squares error minimization method included in the Scipy library \cite{2020SciPy-NMeth}. Such filter has a passband between 0.5 and 40~Hz, using 27 coefficients. The number of coefficients was chosen empirically to obtain good performances with the lowest possible delay. Although such filtering performances might be difficult to obtain in an analog stage positioned before any physical LC-ADC, such preprocessing is used to obtain a realistic comparison between the EB-sampled signal and the dataset: the work in \cite{WFDB_zanoli} shows how EB-ADCs perform by their very nature a filtering process on the signal. Hence, comparing the events defined signal with an unfiltered one would lead to higher error due to bad modeling of unwanted noise as we would compare the results from a de-noised signal (through the action of LC sampling) and a noisy signal. This would mask the error obtained from the event-based sampling reconstruction with the error brought in by the uniform-time ADC used to originally record the dataset.

\textbf{Train-Test subdivision:}
The results of the optimization based methodologies have been obtained with a static Train-Test subdivision: considering the first $N_{hb}$ heartbeats as the train set and the remaining heartbeat as test, without dataset mixing. The levels are learned on the first set and applied to the second one. This was achieved by designing two different splitting granularity: patient-level, and dataset-level subdivision. Both the Bayesian and the random optimization methodology have been studied with both strategies.

The patient-level subdivision splits each patient file in a train set composed of 1'000 heartbeats and keeping the remaining signal as test set (approximately 89'000 heartbeat, depending on each record). The optimization method then uses the train set to learn personalized LC parameters for each recording and apply them to the corresponding test set.

The dataset-level subdivision trains a single LC sampler on a composite train signal, composed of the first 1'000 heartbeats of each record and testing it on all of the test sets.

\textbf{LC-ADC emulation:} Starting from the behavioral definition of LC sampling shown in Eq.\ref{eq:LC_eq}. Then, we include, upon this formulation, the implementation of hysteresis, where each level crossing is indeed constituted of two sub-levels, activated alternatively in Schmitt trigger fashion, where only one sub-level at a time is sensible to the signal crossing, and the signal crossing makes the active level inactive and vice versa. In the optimization-based methods we allow the hysteresis to vary from 0 to $10^{-2}$, such maximum value was experimentally chosen since it is 10\%  of the average ECG signal span in the dataset and we do not expect any noise to be above such threshold, especially after filtering. Moreover, the initial exploration of the methodology showed a significant deterioration of signal reconstruction after this value. The statically defined methods (uniform and logarithmic levels) require a priori estimation of the noise to determine an optimal hysteresis value. Since the signal has been filtered we expect the noise to be negligible. Moreover, an initial exploration of optimization-based methods shows a locally optimal value for such a parameter of 0. Hence, we use a hysteresis of 0 for the statically defined methodologies.

When searching for the crossing of specific values in already recorded signals, the timing question arises: most of the time, such crossing will have happened between two sampling instants. Assuming the Nyquist rate to be respected for these signals, the Sinc interpolation formula could be inverted to obtain the desired timing for a specific signal value. Nonetheless, such an approach does not admit a closed-form solution and would make the computation of the events too expensive for any optimization problem to be considered. However, the timing can be approximated by the inversion of the linear interpolation between two points since this is effectively the solution of the first-order Taylor expansion of the interpolation formula between two points. Given two samples in the uniformly time-sampled signal positioned such that the true signal must pass between one or more levels, we compute the linear equation passing between those points and solve the inverse form for the time corresponding to the levels under analysis.

\textbf{Error estimation:} To effectively show the differences in the analyzed meta-parameter selection methodologies types we define three metrics of particular interest: 1) RMSE per-heartbeat distribution, 2) SRF, and 3) RMSE per event, and compute each of them for LC sampling schemes with an increasing number of levels: 4, 8, 12, and 16.

\textit{RMSE per-heartbeat:} Given the event-based sampled test signal resulting from LC sampling, we linearly interpolate it to obtain a point-to-point comparable signal with the original uniform time sampled signal, here used as ground truth. Then, we subdivide it into single heartbeats: the MIT-BIH dataset is composed of records and annotations marking the timing of each R-peak in the signal (the apex electrical activity of a heartbeat). A heartbeat is then, in this study, the segment of signal between two consecutive R-peaks. 
Then, we compute, for each heartbeat, the RMS error (Eq.~\ref{eq:obj}, with $\lambda=0$). Finally, we show, for every methodology, a box-plot graph describing the main statistics of this metric.

\textit{SRF:} As described in Eq.~\ref{eq:SRF}, this metric is a measurement of the compression ration achieved by LC sampling. We will show, with a line plot, how the SRF changes as we increase the number of levels for each meta-parameters definition methodology.

\textit{RMSE per event:} This combined metric is the result of the RMSE distribution multiplied by the relative SRF. This result, mathematically defined in Eq.~\ref{eq:norm_error}, shows how the error per event is distributed for the different methodologies, giving a complete picture of the performance of the different LC sampling methods.

\subsection{Bayesian optimization sensitivity analysis}
\label{sub-sec:ANOVA_setup}
We opted for the Bayesian optimization implementation in the \texttt{skopt} \cite{skopt} library. To determine the sensitivity of the Bayesian optimization algorithm to its parameters in the current context, we  analyze the fit results of the linear mixed model shown in Eq.\ref{eq:linear_model}. In order to fit this model, we designed a number of experiments such that every corner case of the function is captured. Since we are analyzing five different factors, this leads to $2^5=32$ experiments, each one in a different corner of the domain. Such a experimental design is commonly called a full-factorial design. Although such a design does not take into consideration any central point of the domain, it is usually enough to characterize the linear part of the polynomial approximation of an experimental model.

To define the full-factorial design we need the experiment domain boundaries. A preliminary analysis of the Bayesian optimization algorithm, on a reduced dataset of two recordings and 20'000 heartbeats per recording, resulted in the following set of values for the analyzed factors:

\textbf{Initially sampled points ($I_p$)}: This variable can theoretically vary from two points up to infinity. The sensitivity analysis of this parameter makes it vary from a minimum of 10 points, where we have seen the algorithm starts to be well behaved, to 200, since it appears to have no further improvement beyond this value.

\textbf{Optimally sampled points ($S_p$)}: This variable can varies from one to infinity. However, high values lead to an intractable problem when conditioning the Gaussian process. A tractable range of such variable starts from 20 points, where we have observed the algorithm managing to find some interesting optimal conditions, until 300 points, where the preliminary analysis shows little to no improvements for higher values.

\textbf{Minimum improvement ($\xi$)}: The range of such a variable is intrinsically defined by the magnitude of the results. Too small values of such parameter causes the effective improvement to be random, as the variance of Eq.\ref{eq:utility_f} is effectively higher than the expected improvement. Too high values lead, conversely, to lack of queryable points and a Gaussian process unexplored in areas where the improvement is expected (but not forced) to be low. The magnitude order of the error in Eq.\ref{eq:obj} is $10^{-2}$. Hence, we study this variable from $10^{-3}$ to $5\cdot10^{-2}$ as this range results in significant improvements without unsampled regions of the landscape.

\textbf{Train length($T_l$)}: Since each record is comprised of approximately 90'000 heartbeat, any training set of few hundreds of heartbeats do not sensibly change the relative size of the test set (defined as the remaining heartbeats). Short training sizes allow for faster convergence and a more fine-tuned behaviour on the typical heartbeats observed at the beginning of a record. However, it might discard anomalous behaviors that could be completely misrepresented by the resulting LC parameters. On the contrary, big training sets could allow for more precise outliers representation, at the cost of possibly wrongly assuming the relative frequency of such type of heartbeats, making the resulting LC scheme ill optimized for the more common situations. Hence, a suitable range for such parameter varies from 100 heartbeats (circa one minute of recording) to 1'000 (circa 10 minutes).

\textbf{Lambda ($\lambda$)}: Since the SRF is in the order of $10^{-1}$, and the error has a magnitude order of $10^{-2}$, in these experiments $\lambda$ varies from $5\cdot10^{-2}$, where regularization is a soft conditioning (up to few percent points of the total error), to $1$, where the SRF is as important as the error between the original signal and its reconstruction.

Finally, the sensitivity analysis is performed on an LC sampler with eight levels, as, from the aforementioned preliminary analysis, it results in a sweet spot between the speed of the Bayesian optimization algorithm when applied to this dataset and boundaries, while preserving good results on par with LC acquisition with 12 levels. 




\section{Results}
\label{sec:results}
In this Section, we present three metrics that evaluate the performance of the discussed sampling schemes. Section~\ref{sub-sec:RMSE_res} shows the RMSE-per-beat distribution for each method. Then, in Section \ref{sub-sec:SRF_res} we compare the compression ratio, exemplified by the SRF merit figure. Section \ref{sub-sec:RMSE_dot_SRF_res} presents the per beat distribution of the average error per event. Finally, Section~\ref{sub-sec:ANOVA_res} shows how the selected Bayesian optimization algorithm depends on its meta-pa-rameters, giving the reader an insight on the sensitivity of gradient-free approaches to the level-crossing optimization problem.

In the following, we present the results for the main four methodologies with the optimizations computed for each different patient. Global optimization approaches, in which single solutions are obtained for the entire dataset, achieve marginally lower performance across all experimental settings, but do not provide additional insights regarding the relative benefits and pitfalls of the four investigated LC methodologies.  We report them in Table \ref{tab:full_results_table}, which is discussed in Section \ref{sec:discussion}.

\subsection{Root Mean Square Error distribution}
\label{sub-sec:RMSE_res}
\begin{figure}[ht]
    \centering
    \includegraphics[width=\columnwidth]{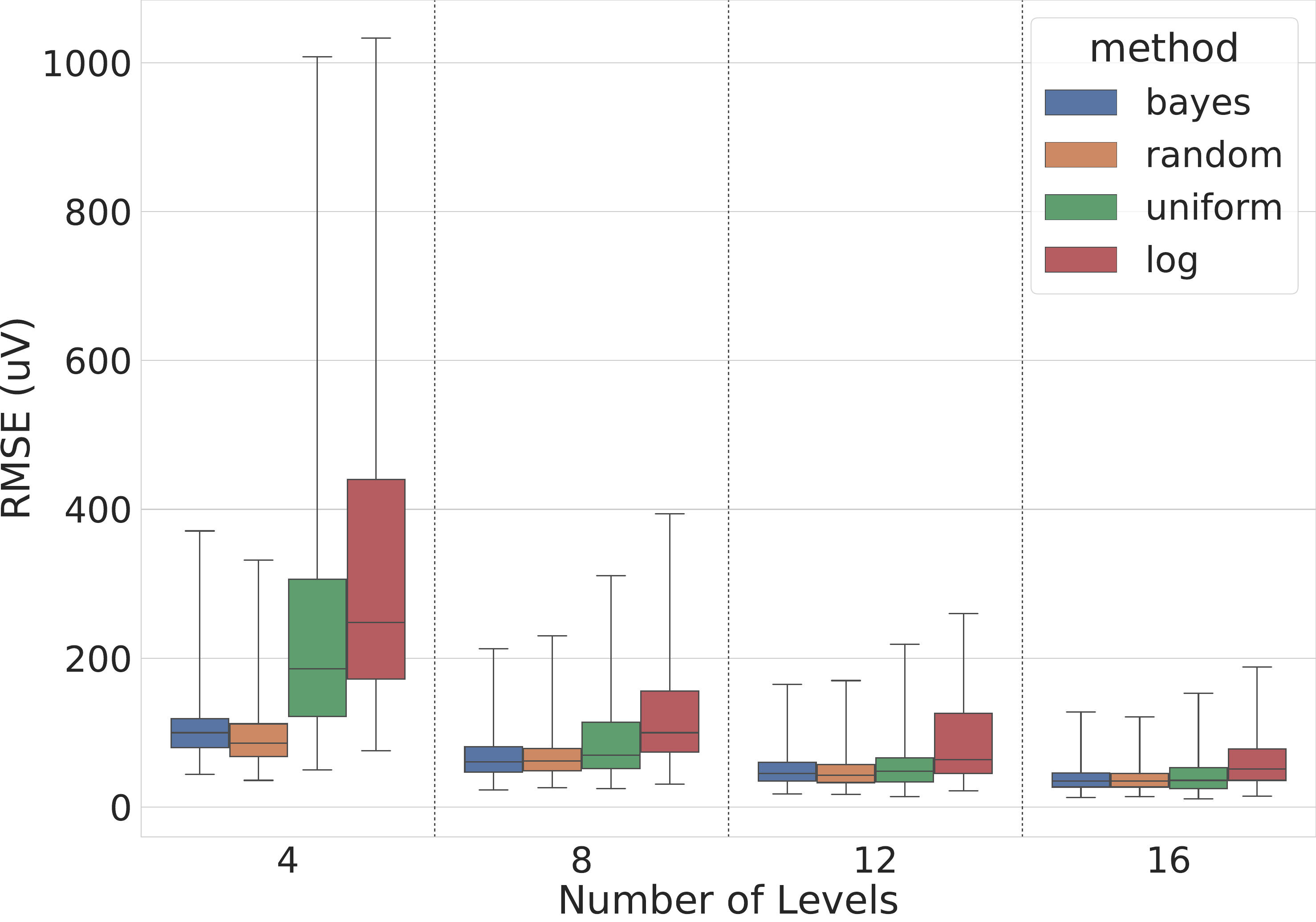}
    \caption{RMSE(mV) box-plot distribution for the principal 4 methodologies for each level analyzed. Center line: Median, top and bottom of each box: $75^{th}$ and $25^{th}$ percentile. Whiskers: 0.5 and 99.5 percentiles.}
    \label{fig:RMSE_dist}
\end{figure}

Figure \ref{fig:RMSE_dist} shows the experimental results relative to the RMS error distribution.

The tabulated view of our results in Table~\ref{tab:RMSE_dist} shows a very similar behaviour for both the gradient-free optimization methods selected(Baysian and random search), while sensibly improving over the statically defined methodologies. In particular, we observe that each optimized LC sampler produces an RMSE of the same order of solutions with uniformly separated levels with four more levels. We notice how level-crossing sampling with 4 levels improves by more than 40\% when applying optimization-based methods, making such potential LC scheme behave comparably to uniform/logarithmic LC schemes \emph{with double the number of levels}.

\begin{table}[ht]
\resizebox{\columnwidth}{!}{%
\begin{tabular}{l|llll}
Levels & Uniform & Logarithmic & Random opt. & Bayesian opt. \\ \hline
4      & 251±196 & 337±231     & 101±107       & 108±96        \\
8      & ~90±63   & 136±95      & ~72±97       & ~71±87         \\
12     & ~56±46   & ~88±58       & ~52±89       & ~54±84         \\
16     & ~43±40   & ~62±39       & ~42±82       & ~42±82        
\end{tabular}
}
\caption{Tabular view of the RMSE main statistics: average and standard deviation for each methodology and number of levels (values in $\mu V$).}
\label{tab:RMSE_dist}
\end{table}

\subsection{Sampling reduction factor comparison}
\label{sub-sec:SRF_res}
\begin{figure}[ht]
    \centering
    \includegraphics[width=\columnwidth]{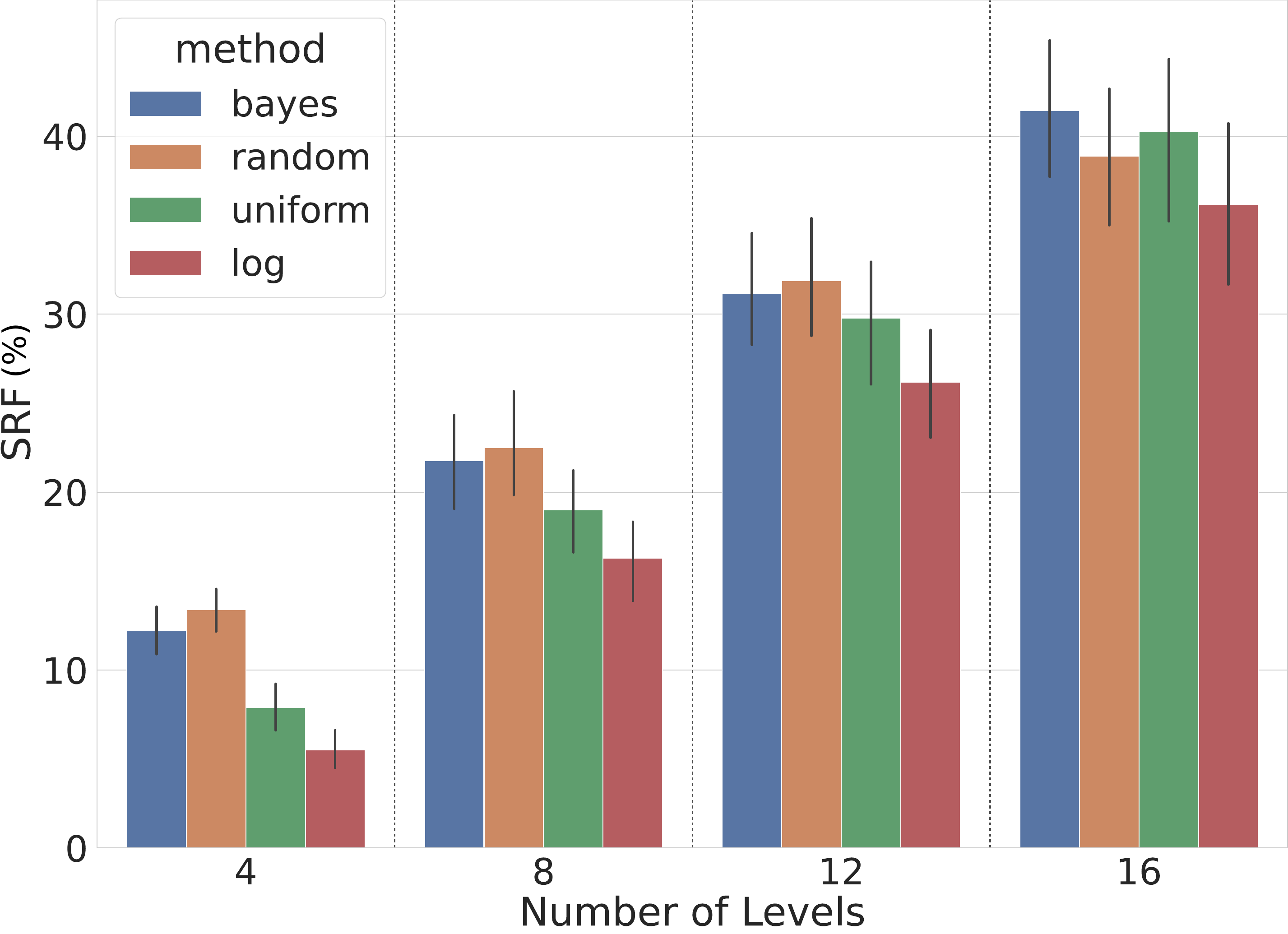}
    \caption{SRF bar-plot average (solid bar) and standard deviation (vertical line atop the bar) for the main 4 methodologies, for each level analyzed.}
    \label{fig:SRF_dist}
\end{figure}

Figure \ref{fig:SRF_dist} shows the SRF distribution between all the patients in the dataset for the different methodologies and number of levels analyzed. 

Table \ref{tab:SRF_dist} numerically explicits the the two main statistics shown in Figure \ref{fig:SRF_dist}. The main takeaway message of these results is that, while the optimization based methodologies result in a smaller overall error, such performance increase is obtained at the price of recording more events. 

\begin{table}[ht]
\resizebox{\columnwidth}{!}{%
\begin{tabular}{l|llll}
Levels & Uniform & Logarithmic & Random opt. & Bayesian opt. \\ \hline
4      & ~7.9±2.8 \%    & ~5.5±2.2 \%        & 13.4±2.6 \%       & 12.2±2.8 \%         \\
8      & 19.0±5.0  \%  & 16.3±5.0 \%       & 22.5±6.3 \%       & 21.8±5.7 \%         \\
12     & 29.8±7.4 \%   & 26.2±7.1 \%       & 31.9±7.1 \%       & 31.1±6.9 \%         \\
16     & 40.3±10 \%    & 36.1±9.6 \%       & 38.9±8.5 \%       & 41.4±8.4 \%        
\end{tabular}%
}
\caption{Tabular view of the SRF main statistics: average and standard deviation for each methodology and each level.}
\label{tab:SRF_dist}
\end{table}

\subsection{Error per event distribution}
\label{sub-sec:RMSE_dot_SRF_res}
\begin{figure}[ht]
    \centering
    \includegraphics[width=\columnwidth]{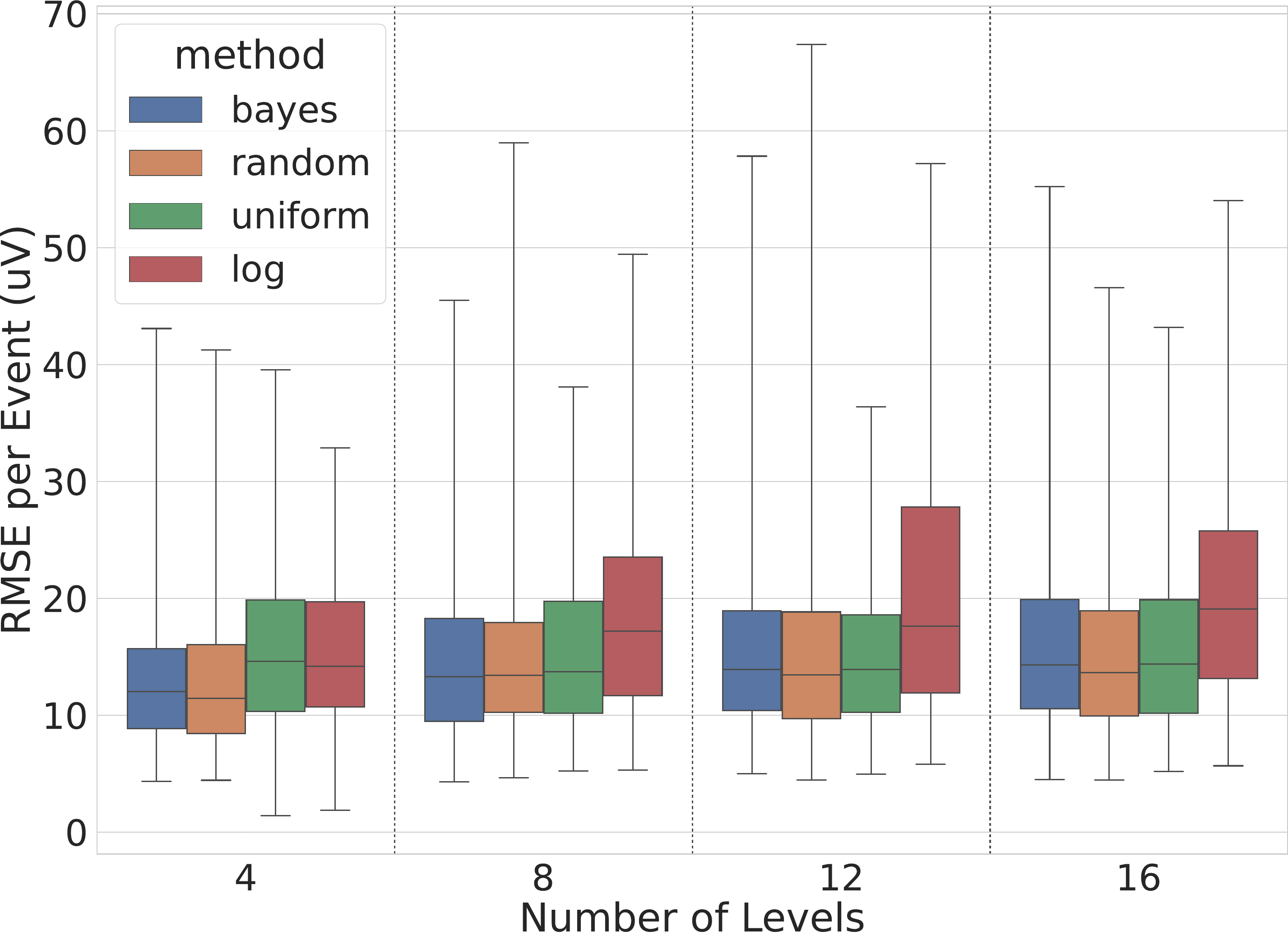}
    \caption{RMSE X SRF box-plot distribution for the four methodologies for each level analyzed. Center line: Median, top and bottom of each box: $75^{th}$ and $25^{th}$ percentile.  Whiskers: 0.5 and 99.5 percentile.}
    \label{fig:RMSE_X_SRF_dist}
\end{figure}

From the two previous results, we can obtain a third composite metric that normalizes the RMSE with respect to the number of events, following eq.~\ref{eq:norm_error}. Such measure intuitively quantifies how much each event contributes to the overall error. Figure~\ref{fig:RMSE_X_SRF_dist}, and the relative numerical representation in Table~\ref{tab:SRF_dot_RMSE_dist} shows how, except for four-level LC sampling solutions, logarithmic level spacing is always outperformed by any other methodology. At the same time, both non-optimized (uniform levels) and gradient-less optimized (random and Bayesian optimization) methodologies are almost indistinguishable. This outcome suggests that there exists a locally optimal limit, when using gradient-less optimization in LC sampling metaparameter search. Solutions that improve the RMSE performance do so by having levels such that the acquired events are not any more significant than other levels configuration, but rather in a bigger number.

\begin{table}[ht]
\resizebox{\columnwidth}{!}{%
\begin{tabular}{l|llll}
Levels & Uniform & Logarithmic & Random opt. & Bayesian opt. \\ \hline
4      & 16±9    & 15±7        & 13±13        & 13±5          \\
8      & 15±7    & 20±11       & 16±23        & 15±7          \\
12     & 15±8    & 21±11       & 17±26        & 16±7          \\
16     & 16±10   & 21±10       & 16±30        & 18±35         
\end{tabular}%
}
\caption{Tabular view of the RMSE times SRF main statistics: average and standard deviation for each methodology and each level (values in $\mu V$).}
\label{tab:SRF_dot_RMSE_dist}
\end{table}

\subsection{Bayesian optimization factor sensitivity}
\label{sub-sec:ANOVA_res}
\begin{figure}[ht]
    \centering
    \includegraphics[width=\columnwidth]{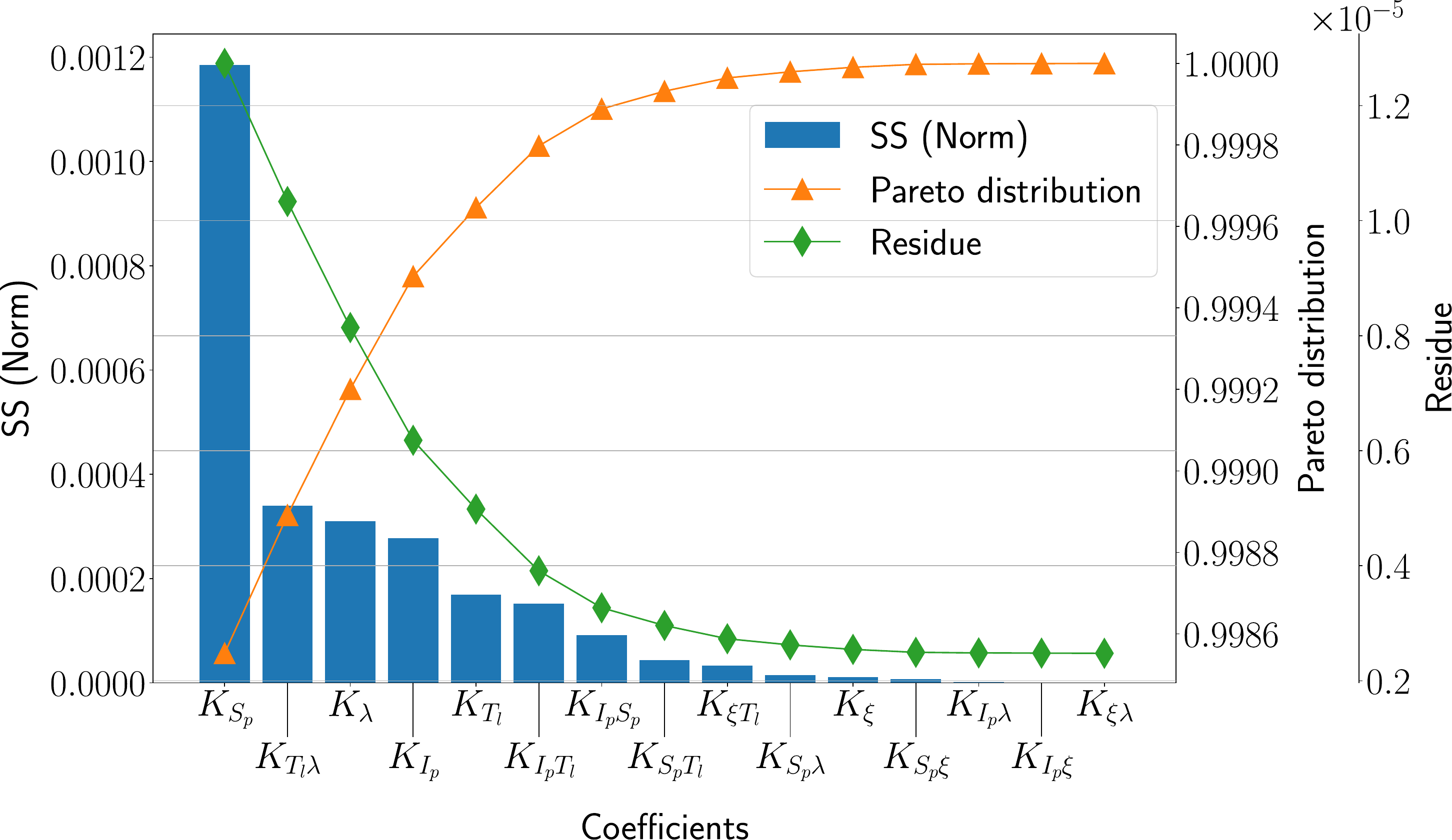}
    \caption{Sensitivity analysis of Bayesian optimization factors. Bar graph: factors sum of square, normalized with respect to the constant factor. Triangle: Pareto distribution of the normalized sum of square. Rhomboids: residue of the model for each subsequent factor.}
    \label{fig:sens_ansys}
\end{figure}

Section \ref{sub-sec:ANOVA_description} highlights how Bayesian optimization is a highly parametric methodology. We are hence interested in studying the impact of such factors on the resulting LC sampling error. We achieve that by finding the best fit for Eq.~\ref{eq:linear_model}, showing how each parameter impacts the obtained RMSE.
Figure~\ref{fig:sens_ansys} summarizes our findings, showing the sum of squares (SS) for each model factor and each two-way combination, normalized with respect to the constant coefficient of the model. We also show the Pareto distribution of the SS with respect to the sum of all factors (including the constant) and the residue reduction brought by each subsequent parameter. First, we can see how the results are mainly explained by the first four factors, namely, the number of Bayesian optimization iterations, the interaction between the length of the train and the lambda regularizer, the lambda regularizer alone, and the initially sampled points, with $K_{sp}$ being the biggest contributing factor by an order of magnitude. However, we notice how the actual magnitudes of such effects are almost negligible when compared to the model constant coefficient: The factor that impacts the most, $K_{sp}$, is just $0.12\%$ the magnitude of the constant, while the model constant accounts for $99.8\%$ of the graphed Pareto distribution. 

These results show how the Bayesian optimization method, applied to our specific problem, is not significantly affected by the meta-parameter particular values and that the only significant factor is the amount of sample points.
The previous findings showing how Bayesian optimization perform on the same level as Random search optimization and the analysis of the error-per-event distribution,  sheds some light on the reason of such a low impact of the model coefficients. After a certain value of RMSE there exist several equally optimal LC parameters, and while more iterations of the Bayesian methodology lead to better results, the basic uncertainty modeling of the sampling function is already enough to obtain one of such numerous optimal LC sampler.

\section{Discussion and Conclusion}
\label{sec:discussion}

In this work, we have studied the performance of level crossing (LC) sampling with four different strategies for levels definition: two statically defined (uniform and logarithmic levels) and two optimization-based methods (random search and Bayesian optimization), focusing on the ability to balance signal fidelity and data compression. We have analyzed long-term ECG signals as a case study, using, as merit figures, the root mean square error (RMSE), the fraction of events relative to the uniform-time sampled signal: the sampling reduction factor (SRF), and error-per-event metrics (i.e., the RMSE normalized with respect to the compression ratio).
Table \ref{tab:full_results_table} shows comprehensive views of the results obtained for all the considered methods and metrics. 

Our results have demonstrated that optimization-based methodologies achieve better signal reconstruction than their static counterparts, with RMSE reductions of up to 40\% compared to uniform levels at equivalent configurations. As an illustrative example, using eight levels, Bayesian optimization yielded an RMSE of 71 ± 87 $\mu$V compared to 90 ± 63 $\mu$V for uniform spacing. Similarly, random optimization performed comparably, with RMSE values of 72 ± 97 $\mu$V at eight levels. However, these optimized methods require more frequent sampling. For the configuration with the same levels, using in particular eight levels, the SRF for Bayesian optimization was 21.8 ± 5.7\%, compared to 19.0 ± 5.0\% for uniform levels, highlighting a trade-off between fidelity and compression.

The error-per-event metric has revealed that optimization methods provide increased signal fidelity while acquiring more samples, and hence achieving better performance by distributing the errors among an increased number of events.

Finally, it is worth highlighting the behavior of four-level LC-Sampling, which, when optimized, manages to behave approximately as well as uniformly spaced LC sampling with double the number of levels.

These findings underscore that the choice of LC level definition must be guided by application requirements. Static configurations, while less optimal, offer simplicity and efficiency for resource-constrained systems. In contrast, Bayesian optimization provides improved accuracy, particularly for scenarios where signal fidelity is critical. Future research should explore adaptive or hybrid approaches that dynamically balance compression and fidelity based on signal characteristics, further advancing the potential of LC sampling in wearable and energy-constrained applications.

\begin{table*}[ht]
{%
\begin{tabular}{c|l|llllll}
\multicolumn{1}{l|}{Metric}                                                                   & Levels & Uniform   & Logarithmic & Random opt$._1$ & Random opt$._2$ & Bayesian opt$._1$ & Bayesian opt$._2$  \\ \hline\hline
\multirow{4}{*}{\begin{tabular}[c]{@{}c@{}}RMSE\\ ($\mu V$) \end{tabular}}                        & 4      & 251±196   & 337±231     & 101±106     & 124±92               & 108±93        & 156±85                 \\
                                                                                              & 8      & 90±63     & 137±96      & 72±97       & 91±65                & 71±92         & 92±68                  \\
                                                                                              & 12     & 56±46     & 88±57       & 52±89       & 73±54                & 54±79         & 78±61                  \\
                                                                                              & 16     & 43±39     & 62±38       & 42±83       & 53±46                & 42±84         & 48±56                  \\ \hline\hline
\multirow{4}{*}{\begin{tabular}[c]{@{}c@{}}SRF\\ (\%) \end{tabular}}                            & 4      & 7.9±2.8 \%  & 5.5±2.2 \%    & 13.4±2.6 \%   & 11.8±2.6 \%            & 12.2±2.8 \%     & 8.5±1.3 \%               \\
                                                                                              & 8      & 19.0±5.0 \% & 16.3±5.0 \%     & 22.5±6.3 \%   & 18.2±3.6 \%            & 21.8±5.7 \%     & 17.6±3.9 \%              \\
                                                                                              & 12     & 29.8±7.4 \% & 26.2±7.1 \%   & 31.9±7.1 \%   & 20.6±4.2 \%            & 31.1±6.9 \%     & 20.9±4.5 \%              \\
                                                                                              & 16     & 40.3±10.0\% & 36.2±9.6 \%   & 38.9±8.5 \%   & 30.2±6.4 \%            & 41.4±8.4 \%     & 36.6±7.0 \%              \\ \hline\hline
\multirow{4}{*}{\begin{tabular}[c]{@{}c@{}}RMSE \\ $\cdot$ \\ SRF\\ ($\mu V$) \end{tabular}}  & 4      & 16±9      & 15±7        & 13±13       & 15±10                & 13±11         & 13±7                   \\
                                                                                              & 8      & 15±7      & 20±11       & 16±23       & 17±11                & 15±19         & 16±11                  \\
                                                                                              & 12     & 15±8      & 21±11       & 17±27       & 15±10                & 17±24         & 17±11                  \\
                                                                                              & 16     & 16±10     & 21±10       & 16±30       & 16±12                & 18±36         & 18±18                 
\end{tabular}
}
\caption{Tabular view of all the analyzed metrics: RMSE, SRF, and  RMSE times SRF main statistics: average and standard deviation for each methodology and each level. The sub-script (1) refer to patient-level optimization. Sub-script (2) refer to mixed dataset-level optimization. RMSE and RMS X SRF ($R~\cdot~S$) in $\mu V$. SRF expressed as a percentage}
\label{tab:full_results_table}
\end{table*}

Future works in this field should address two relevant limitations: the relatively small but already complex search space and the homogeneity of the analyzed signal. While this study achieves good performance, no arrhythmia conditions were considered. Moreover, a local hysteresis for each individual level was considered but quickly discarded due to the doubling in search space size, making the Bayesian optimization methodology unfeasible. Finally, an adaptive level learning methodology, where a new set of levels is learned whenever event distributions vary, has the potential to increase the achieved performance at the price of repeating expensive optimization cycles. Hence, the exploration of more efficient gradient-free approaches is essential for optimal and local LC sampling meta-parameter tuning.

\vspace{-2pt}
\section*{Acknowledgements}

This work was supported in part by the Swiss State Secretariat for Education, Research, and Innovation (SERI) through the SwissChips research project, and the EC H2020 DIGIPREDICT Project (GA no. 101017915), and the ACCESS—AI Chip Center for Emerging Smart Systems, sponsored by InnoHK funding. T.T. is supported by the grant RYC2021-032853-I funded by MCIN/AEI/ 10.13039/501100011033 and by the European Union NextGenerationEU/PRTR.


 \bibliographystyle{elsarticle-num} 
 \bibliography{references}

\begin{thebibliography}{10}
\expandafter\ifx\csname url\endcsname\relax
  \def\url#1{\texttt{#1}}\fi
\expandafter\ifx\csname urlprefix\endcsname\relax\def\urlprefix{URL }\fi
\expandafter\ifx\csname href\endcsname\relax
  \def\href#1#2{#2} \def\path#1{#1}\fi

\bibitem{cardio_diseases_leading_death_cause}
W.~H. Organization, {W.H.O.} data on cardiovascular diseases (2019).

\bibitem{Wei19}
L.~{Wei}, {et.al.}, 13.3 a 7mb stt-mram in 22ffl finfet technology with 4ns read sensing time at 0.9v using write-verify-write scheme and offset-cancellation sensing technique, in: 2019 IEEE International Solid- State Circuits Conference - (ISSCC), 2019, pp. 214--216.
\newblock \href {https://doi.org/10.1109/ISSCC.2019.8662444} {\path{doi:10.1109/ISSCC.2019.8662444}}.

\bibitem{Abadal20}
S.~{Abadal}, C.~{Han}, J.~M. {Jornet}, Wave propagation and channel modeling in chip-scale wireless communications: A survey from millimeter-wave to terahertz and optics, IEEE Access 8 (2020) 278--293.
\newblock \href {https://doi.org/10.1109/ACCESS.2019.2961849} {\path{doi:10.1109/ACCESS.2019.2961849}}.

\bibitem{Panades20}
I.~{Miro-Panades}, {et.al.}, {SamurAI: A 1.7MOPS-36GOPS Adaptive Versatile IoT Node with 15,000x Peak-to-Idle Power Reduction, 207ns Wake-Up Time and 1.3TOPS/W ML Efficiency}, in: 2020 IEEE Symposium on VLSI Circuits, 2020, pp. 1--2.
\newblock \href {https://doi.org/10.1109/VLSICircuits18222.2020.9163000} {\path{doi:10.1109/VLSICircuits18222.2020.9163000}}.

\bibitem{Pullini18}
A.~{Pullini}, D.~{Rossi}, I.~{Loi}, A.~{Di Mauro}, L.~{Benini}, {Mr. Wolf: A 1 GFLOP/s Energy-Proportional Parallel Ultra Low Power SoC for IOT Edge Processing}, in: ESSCIRC 2018 - IEEE 44th European Solid State Circuits Conference (ESSCIRC), 2018, pp. 274--277.
\newblock \href {https://doi.org/10.1109/ESSCIRC.2018.8494247} {\path{doi:10.1109/ESSCIRC.2018.8494247}}.

\bibitem{rincon2011development}
F.~Rinc{\'o}n, J.~Recas, N.~Khaled, D.~Atienza, Development and evaluation of multilead wavelet-based ecg delineation algorithms for embedded wireless sensor nodes, IEEE TITB 15~(6) (2011) 854--863.

\bibitem{Shannon1949}
C.~Shannon, Communication in the presence of noise, Proceedings of the {IRE} 37~(1) (1949) 10--21.
\newblock \href {https://doi.org/10.1109/jrproc.1949.232969} {\path{doi:10.1109/jrproc.1949.232969}}.

\bibitem{level_crossing}
G.~Rovere, S.~Fateh, L.~Benini, A 2.1 {$\mu$}w event-driven wake-up circuit based on a level-crossing adc for pattern recognition in healthcare, in: 2017 IEEE Biomedical Circuits and Systems Conference (BioCAS), 2017, pp. 1--4.
\newblock \href {https://doi.org/10.1109/BIOCAS.2017.8325145} {\path{doi:10.1109/BIOCAS.2017.8325145}}.

\bibitem{analog_level_crossing}
N.~Sayiner, H.~Sorensen, T.~Viswanathan, A level-crossing sampling scheme for a/d conversion, IEEE Transactions on Circuits and Systems II: Analog and Digital Signal Processing 43~(4) (1996) 335--339.
\newblock \href {https://doi.org/10.1109/82.488288} {\path{doi:10.1109/82.488288}}.

\bibitem{polygonal_approximator}
S.~Zanoli, F.~Ponzina, T.~Teijeiro, A.~Levisse, D.~Atienza, An error-based approximation sensing circuit for event-triggered, low power wearable sensors (2021).
\newblock \href {https://doi.org/10.48550/ARXIV.2106.13545} {\path{doi:10.48550/ARXIV.2106.13545}}.

\bibitem{log_lvls}
S.~Sirimasakul, A.~Thanachayanont, A logarithmic level-crossing adc with fixed comparison window, in: 2022 19th International Conference on Electrical Engineering/Electronics, Computer, Telecommunications and Information Technology (ECTI-CON), 2022, pp. 1--4.
\newblock \href {https://doi.org/10.1109/ECTI-CON54298.2022.9795458} {\path{doi:10.1109/ECTI-CON54298.2022.9795458}}.

\bibitem{jax2018github}
J.~B. et~al., \href{http://github.com/jax-ml/jax}{{JAX}: composable transformations of {P}ython+{N}um{P}y programs} (2018).
\newline\urlprefix\url{http://github.com/jax-ml/jax}

\bibitem{non_uniform_sampling_reconstruction_conditions}
A.~Aldroubi, K.~Gröchenig, Nonuniform sampling and reconstruction in shift-invariant spaces, SIAM Review 43~(4) (2001) 585--620.
\newblock \href {https://doi.org/10.1137/S0036144501386986} {\path{doi:10.1137/S0036144501386986}}.

\bibitem{sub_nyquist_optimal}
R.~Venkataramani, Y.~Bresler, Optimal sub-nyquist nonuniform sampling and reconstruction for multiband signals, IEEE Transactions on Signal Processing 49~(10) (2001) 2301--2313.
\newblock \href {https://doi.org/10.1109/78.950786} {\path{doi:10.1109/78.950786}}.

\bibitem{compressed_sensing}
M.~Taghouti, Chapter 10 - compressed sensing, in: F.~H. Fitzek, F.~Granelli, P.~Seeling (Eds.), Computing in Communication Networks, Academic Press, 2020, pp. 197--215.
\newblock \href {https://doi.org/10.1016/B978-0-12-820488-7.00023-2} {\path{doi:10.1016/B978-0-12-820488-7.00023-2}}.

\bibitem{LC_lvls_in_SRF_function}
M.~Miskowicz, Send-on-delta concept: An event-based data reporting strategy, Sensors 6~(1) (2006) 49--63.
\newblock \href {https://doi.org/10.3390/s6010049} {\path{doi:10.3390/s6010049}}.

\bibitem{Average_Samp_freq_LC}
M.~Ji, K.~M. Chugg, Analysis of the average sampling frequency for level crossing analog-to-digital converters, in: MILCOM 2021 - 2021 IEEE Military Communications Conference (MILCOM), 2021, pp. 121--126.
\newblock \href {https://doi.org/10.1109/MILCOM52596.2021.9653065} {\path{doi:10.1109/MILCOM52596.2021.9653065}}.

\bibitem{EB-PAS-analysis}
M.~Miskowicz, Asymptotic effectiveness of the event-based sampling according to the integral criterion, Sensors 7~(1) (2007) 16--37.
\newblock \href {https://doi.org/10.3390/s7010016} {\path{doi:10.3390/s7010016}}.

\bibitem{EB_for_control}
M.~Miśkowicz, Event-based sampling strategies in networked control systems, in: 2014 10th IEEE Workshop on Factory Communication Systems (WFCS 2014), 2014, pp. 1--10.
\newblock \href {https://doi.org/10.1109/WFCS.2014.6837603} {\path{doi:10.1109/WFCS.2014.6837603}}.

\bibitem{uniform_lvls_control}
K.~{Johan Åström}, B.~Bernhardsson, Comparison of periodic and event based sampling for first-order stochastic systems, IFAC Proceedings Volumes 32~(2) (1999) 5006--5011, 14th IFAC World Congress 1999, Beijing, Chia, 5-9 July.
\newblock \href {https://doi.org/10.1016/S1474-6670(17)56852-4} {\path{doi:10.1016/S1474-6670(17)56852-4}}.

\bibitem{many_EB-samp_for_control}
V.~Vasyutynskyy, K.~Kabitzsch, Towards comparison of deadband sampling types, in: 2007 IEEE International Symposium on Industrial Electronics, 2007, pp. 2899--2904.
\newblock \href {https://doi.org/10.1109/ISIE.2007.4375074} {\path{doi:10.1109/ISIE.2007.4375074}}.

\bibitem{pdf_based_LCS}
H.~Alasti, Non-uniform-level crossing sampling for efficient sensing of temporally sparse signals, IET Wireless Sensor Systems 4~(1) (2014) 27--34.
\newblock \href {https://doi.org/10.1049/iet-wss.2013.0010} {\path{doi:10.1049/iet-wss.2013.0010}}.

\bibitem{adaptive_LC_recon}
S.~Senay, L.~F. Chaparro, M.~Sun, R.~J. Sclabassi, Adaptive level-crossing sampling and reconstruction, in: 2010 18th European Signal Processing Conference, 2010, pp. 1296--1300.

\bibitem{energy_intensive_cool_adaptive_LC}
W.~Tang, A.~Osman, D.~Kim, B.~Goldstein, C.~Huang, B.~Martini, V.~A. Pieribone, E.~Culurciello, Continuous time level crossing sampling adc for bio-potential recording systems, IEEE Transactions on Circuits and Systems I: Regular Papers 60~(6) (2013) 1407--1418.
\newblock \href {https://doi.org/10.1109/TCSI.2012.2220464} {\path{doi:10.1109/TCSI.2012.2220464}}.

\bibitem{WFDB_zanoli}
S.~Zanoli, G.~Ansaloni, T.~Teijeiro, D.~Atienza, Event-based sampled ecg morphology reconstruction through self-similarity, Computer Methods and Programs in Biomedicine 240 (2023) 107712.
\newblock \href {https://doi.org/10.1016/j.cmpb.2023.107712} {\path{doi:10.1016/j.cmpb.2023.107712}}.

\bibitem{Byesian_opt_book}
J.~Mockus, Bayesian Approach to Global Optimization, Springer Dordrecht, 1989.
\newblock \href {https://doi.org/10.1007/978-94-009-0909-0} {\path{doi:10.1007/978-94-009-0909-0}}.

\bibitem{bayesian_opt_review}
X.~Wang, Y.~Jin, S.~Schmitt, M.~Olhofer, Recent advances in bayesian optimization, ACM Comput. Surv. 55~(13s) (Jul. 2023).
\newblock \href {https://doi.org/10.1145/3582078} {\path{doi:10.1145/3582078}}.

\bibitem{fractal_ECG_reconstruction}
A.~Ibaida, D.~Al-Shammary, I.~Khalil, Cloud enabled fractal based ecg compression in wireless body sensor networks, Future Generation Computer Systems 35 (2014) 91--101, special Section: Integration of Cloud Computing and Body Sensor Networks; Guest Editors: Giancarlo Fortino and Mukaddim Pathan.
\newblock \href {https://doi.org/10.1016/j.future.2013.12.025} {\path{doi:10.1016/j.future.2013.12.025}}.

\bibitem{physioToolkit}
A.~Goldberger, {et.al.}, Physiobank, physiotoolkit, and physionet: Components of a new research resource for complex physiologic signals (2000).
\newblock \href {https://doi.org/10.13026/C2F30} {\path{doi:10.13026/C2F30}}.

\bibitem{HR_BW}
J.~J. Bailey, A.~S. Berson, A.~Garson, L.~G. Horan, P.~W. Macfarlane, D.~W. Mortara, C.~Zywietz, Recommendations for standardization and specifications in automated electrocardiography: bandwidth and digital signal processing. a report for health professionals by an ad hoc writing group of the committee on electrocardiography and cardiac electrophysiology of the council on clinical cardiology, american heart association., Circulation 81~(2) (1990) 730--739.
\newblock \href {https://doi.org/10.1161/01.CIR.81.2.730} {\path{doi:10.1161/01.CIR.81.2.730}}.

\bibitem{2020SciPy-NMeth}
P.~Virtanen, G.~et~al., {{SciPy} 1.0: Fundamental Algorithms for Scientific Computing in Python}, Nature Methods 17 (2020) 261--272.
\newblock \href {https://doi.org/10.1038/s41592-019-0686-2} {\path{doi:10.1038/s41592-019-0686-2}}.

\bibitem{skopt}
T.~H. et~al., scikit-optimize/scikit-optimize: v0.5.2 (mar 2018).
\newblock \href {https://doi.org/10.5281/zenodo.1207017} {\path{doi:10.5281/zenodo.1207017}}.

\end{thebibliography}





\end{document}